\documentclass[aps, pra, 10pt, twocolumn]{revtex4-1}

\usepackage{graphicx} % Include figure files
\usepackage{hyperref} % Hyperlinks
\usepackage{mathptmx} % Use Times for math and text
\usepackage{xcolor}   % More colors!
\usepackage{bm}       % Bold Greek symbols
\usepackage{physics}  % Bra/kets amongst other things.

\hypersetup{
    colorlinks=true,
    linkcolor=blue,
    citecolor=blue,
    filecolor=blue,
    urlcolor=blue
}

\newcommand{\NIST}{Sensor Science Division, National Institute of Standards and Technology, Gaithersburg, Maryland 20899, USA}

\graphicspath{{./}{./figs/}}

\begin{document}

\title{Simulations of a frequency-chirped magneto-optical trap of MgF}
\author{Kayla J. Rodriguez}
\altaffiliation{Present address: Joint Quantum Institute, University of Maryland, College Park, MD 20742, USA}
\email{krodrig8@umd.edu}
\affiliation{\NIST}
\author{Nickolas H. Pilgram}
\affiliation{\NIST}
\author{Daniel S. Barker}
\affiliation{\NIST}
\author{Stephen P. Eckel}
\email{stephen.eckel@nist.gov}
\affiliation{\NIST}
\author{Eric B. Norrgard}
\email{eric.norrgard@nist.gov}
\affiliation{\NIST}

\date{\today}
\begin{abstract}
We simulate the capture process of MgF molecules into a frequency-chirped molecular MOT.
Our calculations show that by chirping the frequency, the MOT capture velocity is increased by about of factor of 4 to 80~m/s, allowing for direct loading from a two-stage cryogenic buffer gas beam source.
Moreover, we simulate the effect of this frequency chirp for molecules already present in the MOT.
We find that the MOT should be stable with little to no molecule loss. The chirped MOT should thus allow loading of multiple molecule pulses to increase the number of trapped molecules.
\end{abstract}

\maketitle
\section{Introduction}
All molecular magneto-optical traps (MOTs) produced to-date \cite{McCarron2018b, Tarbutt2018,Norrgard2016,Burau2023,andereggRadioFrequencyMagnetoOptical2017,williamsCharacteristicsMagnetoopticalTrap2017,luMolecularLaserCooling2022a,vilasMagnetoopticalTrappingSubDoppler2022} have been loaded from a laser-slowed cryogenic buffer gas beam (CBGB).  
Laser slowing is necessitated by the mismatch in velocity scales: the single-stage CBGB source typically produces molecular beams with peak velocities of over 100~m/s \cite{Hutzler2012,Truppe2018} while the typical capture velocity of the MOT is of the order of 10~m/s \cite{williamsCharacteristicsMagnetoopticalTrap2017,TarbuttII2015}. Two-stage CBGB sources are capable of producing slower beams with mean velocities approaching 60~m/s, but still larger than the typical capture velocity of molecular MOTs \cite{Hutzler2012, Lu2011, Hemmerling2014,pattersonBrightGuidedMolecular2007}.

In principle, direct loading of molecular MOTs is possible if the MOT laser beam  is larger than the stopping distance for a incident molecule.  Consider a laser cooling scheme with $n_g$ ground states and $n_e$ excited states.
The maximum possible deceleration is $a_{\rm{max}}=h \Gamma n_e/((n_e+n_g) m  \lambda)$~\cite{Kloter2008,Norrgard2016}, so large deceleration is possible in molecules with low mass $m$, fast radiative decay rate $\Gamma$, and short wavelength $\lambda$.
In order to maintain a large deceleration, direct MOT loading further requires that within the spatial extent of the MOT laser beams, the trapping laser frequency is nearly resonant with the range of Doppler shifts corresponding to velocities between the initial molecular beam velocity and rest.
Typically, this requirement is fulfilled by the MOT's spherical quadrupole magnetic field, which provides a range of Zeeman shifts spanning the requesite range of Doppler shifts.
However, as we shall show, the small $g$ factor of the A$^2\Pi_{1/2}$ excited state typically used for laser cooling alkaline-earth fluoride molecules provides insufficient variation in the Zeeman shift to maintain a resonant interaction over the entire stopping distance.
Therefore, molecular structure, not MOT geometry, generally limits the capture velocity to around 10~m/s.

Absent a substantial excited state $g$-factor, it is possible to engineer a temporally varying laser frequency such that resonant deceleration is maintained as molecules are slowed to a stop.  
In this work, we simulate such a ``chirped MOT'' and show that capture velocities up to roughly 100~m/s are possible with realistic experimental parameters.
We focus on MgF, which has been extensively studied in single-stage CBGBs as a candidate laser coolable molecule \cite{Gu2022,Doppelbauer2022, Norrgard2023} but has not yet been slowed or trapped.
MgF is a good test case because of its relatively large recoil velocity (2.6\, cm/s) and large radiative decay rate ($\Gamma\,=\,131.6(1.4)$\,s$^{-1}$) \cite{Norrgard2023}.
The large capture velocity of the MgF chirped MOT is sufficient to capture nearly all molecules from a two-stage CBGB source (or from a one stage CBGB source with modest laser slowing).
Moreover, we show that trapped molecules in the MOT are retained during a subsequent frequency chirp, thus allowing multiple successive molecular beam pulses to be captured by the MOT.
This result contrasts with typical chirped slowing techniques, which use a single slowing beam that intersects the MOT and causes molecule loss during its frequency chirp.

The concept of chirped laser slowing was proposed in Ref.~\cite{LETOKHOV1976} and utilized  in some of the earliest atomic laser cooling experiments \cite{Prodan1984,Ertmer1985,Chu1985,Raab1987}.  Frequency-chirped MOTs are a common feature in alkaline-earth laser-cooling experiments because of the similar mismatch between capture velocity and velocity of the source~\cite{Katori1999, Kuwamoto1999, Curtis2003}.
In the case of Sr, the source is typically a ``blue'' MOT, operating on the $^1{\rm S}_0\rightarrow\, ^1{\rm P}_1$ transition at 461~nm.
Sr atoms are generally cooled to root-mean-square velocities on the order of 1~m/s, well above the 5~mm/s molasses capture velocity of the Sr ``red'' MOT, which operates on the $^1{\rm S}_0\rightarrow\, ^3{\rm P}_1$ intercombination transition~\cite{Katori1999, Vogel1999, Xu2003, Nagel2008a}.
To increase the capture velocity, the frequency of the ``red'' MOT light is modulated from $\Delta/\Gamma \approx -200$ to $\Delta/\Gamma\approx-10$~\cite{Katori1999, Muniz2018, Snigirev2019}.
This extends the capture velocity to on the order of 1~m/s. %\DSB{double check this one}.\SPE{Must be true, otherwise, it wouldn't work.} \EBN{Maybe use a difference citation?  In the current ciation, it is more like a broadband MOT than a chirped MOT}\DSB{I'll find a better one, turns out that the early adopters didn't do chirps like all the later MOTs did}

Our discussion is organized as follows:
Section~\ref{sec:model} describes our MOT geometry, level structure and molecular Hamiltonian, and the details of the calculations.
%We use the package {\tt pylcp}, a python package for laser cooling~\cite{Eckel2022}, allowing us to integrate complicated geometries and level structures into laser cooling calculations.
Section~\ref{sec:static_MOT} details the properties of static MOTs and their respective capture processes, using a MOT of $^{87}$Rb as a prototypical example.
Section~\ref{sec:chirped_MOT} discusses our proposed frequency chirped MOT, and shows that it can increase the capture velocity by almost a factor of 4.
The results of section~\ref{sec:stability} reveal that our MOT should be stable against the chirp, enabling multi-pulse loading.
Finally, we conclude in Sec.~\ref{sec:conclusion}.

\section{Model}

\label{sec:model}
\begin{figure}
    \centering
    \includegraphics[width=\columnwidth]{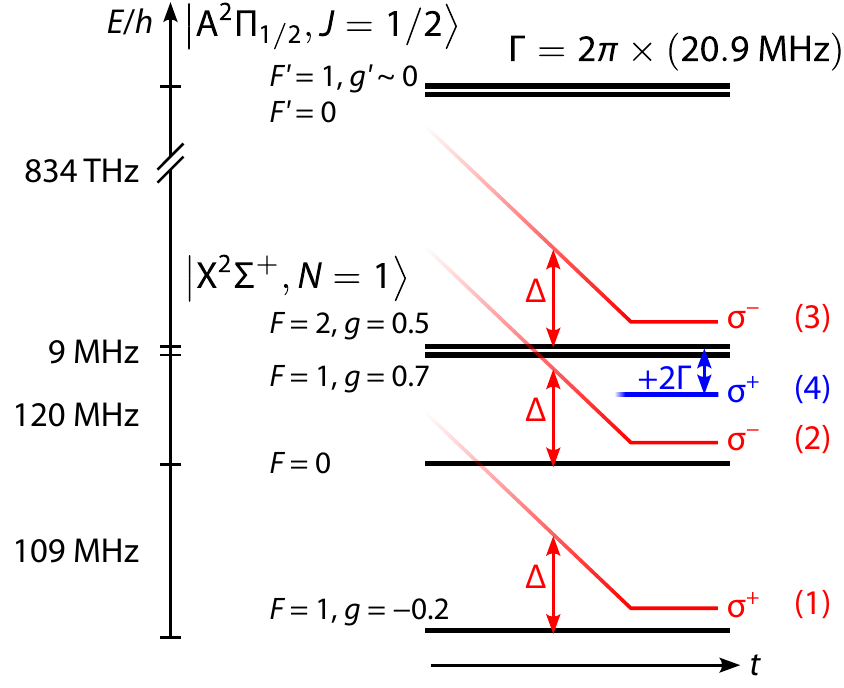}
    \caption{
    Level diagram and laser cooling scheme for a MgF chirped MOT. 
    The $\left|{\rm X}^2\Sigma^+, N=1\right>\rightarrow \left|{\rm A}^2\Pi_{1/2}, J^\prime=1/2\right>$ transition spin-rotation/hyperfine levels are shown with energy spacing $E/h$.
    Four laser frequencies labeled (1)--(4) address the transitions with polarizations $\sigma^\pm$.
    The respective detunings $\Delta$ are shown schematically to vary with time $t$.
    Once frequency components (1)-(3) reach their final value, frequency (4) is added.
    }
    \label{fig:laser_cool_trans}
\end{figure}

We model a six-beam molecular MOT, incorporating the relevant MgF level structure, multiple frequency components in the MOT beams, and changing laser detuning and intensity with time. Molecules enter the MOT along the $x$ axis with a longitudinal velocity $v$ and a much smaller transverse velocity. The magnetic field gradient $\mathbf{B} = B' (-x\hat{x}/2 - y\hat{y}/2 +z\hat{z})$ has its strong axial gradient along $z$. Six laser beams propagate along $\pm x'$, $\pm y'$ and $\pm z$ directions, where the $x'$ and $y'$ axes are rotated from the $x$ and $y$ axes about $z$ by 45$^\circ$. Because the MOT beams enter at 45$^\circ$ with respect to the molecular beam, longitudinal slowing in a chirped MOT should not be substantially different for molecules with a small transverse velocity component. Hence, we simulate motion only along the $\hat{x}$ axis. In our simulations, we use both infinite plane wave beams and elliptical Gaussian beams, depending on the situation. For the latter, the beams with $\mathbf{k}$ in the $x$-$y$ plane of the MOT have a $1/e^2$ radius parallel to the $x$-$y$ plane of $w_{xy}$ and a $1/e^2$ radius along $\hat{z}$ of $w_z$. Likewise, the beams with $\hat{k}$ along the $z$ axis have $1/e^2$ radius of $w_{xy}$ along $x$ and a $1/e^2$ radius of $w_z$ along $y$. All six beams are assumed to have equal peak intensity.

\begin{figure*}
    \centering
    \includegraphics{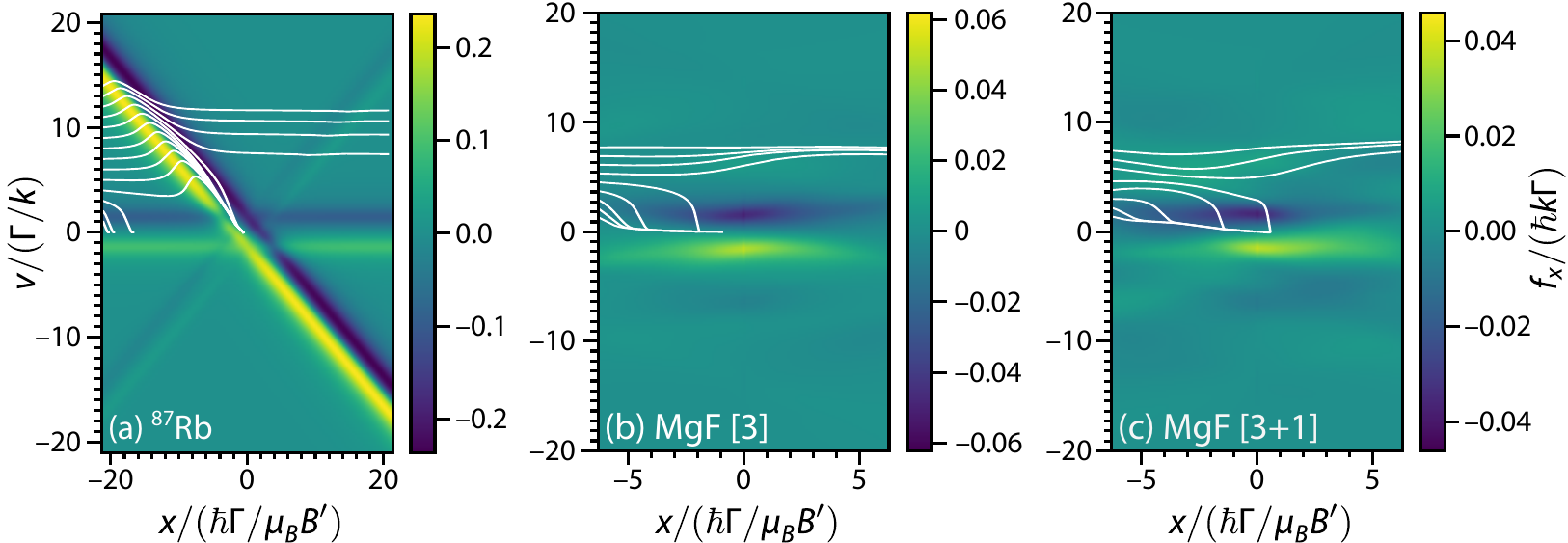}
    \caption{
    Calculated normalized force $f/(\hbar k \Gamma)$ vs. Zeeman-detuning normalized position $x/(\hbar \Gamma/\mu_B B')$ and normalized velocity $v/(\Gamma/k)$, where $k$ is the wavevector of the light, $\Gamma$ is the excited state decay rate, $\mu_B$ is the Bohr magneton, and $B'$ is the magnetic field gradient, in a MOT with infinite plane wave beams for (a) $^{87}$Rb using two frequencies components both with $\Delta_0=-\Gamma$, (b) MgF using three frequency components with $\Delta_m=-\Gamma$ [3], and (c) MgF using three frequency components with $\Delta_m=-\Gamma$ and a fourth with $\Delta_4=+2\Gamma$ [3+1] (see Fig.~\ref{fig:laser_cool_trans}).
    Note the differences in the color scales between the panels.
    The white curves overlaid on the plots show the calculated trajectories using $B'=20$~G/cm.
    }
    \label{fig:explainer_fig}
\end{figure*}

Our model molecular Hamiltonian is computed using parameters of the MgF $\ket{\text{X}^2\Sigma^+, v=0; N=1} \rightarrow \ket{\text{A}^2\Pi_{1/2}, v^\prime=0; J^\prime=1/2}$ laser cooling transition \cite{Doppelbauer2022,xuDeterminationNormalState2019}, with relevant parameters shown in Fig.~\ref{fig:explainer_fig}. For this transition, $\Gamma = 2\pi\times [20.9(2)\mbox{ MHz}]$~\cite{Norrgard2023}, $\omega \approx 2\pi\times (834.3\mbox{ THz})$, and the  effective two-level saturation intensity is $I_{\rm sat}= \hbar \omega^3 \Gamma/(12 \pi c^2)\approx 60$ mW/cm$^2$. Higher vibrational levels $v\ge1$ are ignored in our model; assuming  $v=1$ is repumped on the $\ket{\text{X}^2\Sigma^+, v=1;N=1}\rightarrow \ket{\text{B}^2\Sigma^+, v^\prime=0;N^\prime=0}$ transition, this approximation should only affect the computed capture velocity at the percent level because decays to $v \ge 1$ occur with roughly 3\,\% probability~\cite{Norrgard2023}. The $\ket{\text{X}^2\Sigma^+, v=0; N=1}$ ground state is split into a manifold of four levels by the combinination of spin-rotation and hyperfine interactions.

The effective Hamiltonian is computed in a basis comprised of the 16 Zeeman sublevels of the $\ket{\text{X}^2\Sigma^+, v=0; N=1}$ and $ \ket{\text{A}^2\Pi_{1/2}, v^\prime=0; J^\prime=1/2}$ states. This Hamiltonian accounts for the ground state spin-rotation and dipolar hyperfine interactions, all relevant Zeeman interactions, and the coupling between the states due to the laser fields. Because we are only considering transitions between the Zeeman sublevels of the single $N=1$ rotational level of the $\text{X}^2\Sigma^+$ state and a single $\Lambda$-doublet component of the $J^\prime=1/2$ level of the $\text{A}^2\Pi_{1/2}$ state, the effects of the rotational and $\Lambda$-doubling interactions are neglected. Additionally, mixing with states outside of this manifold due to the Zeeman interaction are negligible. The relevant spectroscopic parameters of MgF can be found in Ref. \cite{Doppelbauer2022,andersonMillimeterWaveSpectroscopy1994} 

The $g$-factor of the $\ket{\text{A}^2\Pi_{1/2}, v^\prime=0; J^\prime=1/2}$ state is nearly zero.
%This state is best described by Hund's case a \cite{Herzberg1950}.  
The effective Zeeman Hamiltonian of a $^2\Pi_{1/2}$ state can be modeled by six distinct magnetic interactions $H_1, \dots , H_6$ plus a nuclear spin Hamiltonian $H_{7}^{(i)}$ for each nucleus $i$ possessing a spin (here, the subscripts correspond to terms of Eq.\,(17) in Ref. \cite{Brown1978}).
Typically, the Zeeman interactions are dominated by the electron spin Zeeman Hamiltonian $H_1 = g_S\mu_B \textbf{B}\cdot \textbf{S}$  and the electron orbital angular momentum Zeeman Hamiltonian $H_2 = g^\prime_L \mu_B  \textbf{B}\cdot \textbf{L}$, so that $g$ is proportional to $ g^\prime_L \Lambda + g_S \Sigma $.  Here, $g_S \approx 2.002$ is the electron $g$-factor corrected for relativistic effects, $g^\prime_L \approx 1$ is the orbital $g$-factor corrected for relativistic effects, and the prime indicates a small additional correction to account for adiabatic effects \cite{Brown1978, Brown2003}.
In $^2\Pi_{1/2}$ states, these terms nearly cancel: $g_L^\prime \Lambda + g_S \Sigma \approx 0.002$. 
In heavier systems which are isoelectronic to MgF (e.g.\ CaF, SrF, and YO), the effective $\text{A}^2\Pi_{1/2}$ $g$-factor is still of order $\vert g \vert\sim 0.1$  \cite{TarbuttII2015}.  
This is because the Zeeman interaction in these systems is dominated by  two parity-dependent Zeeman interactions $H_5$ and $H_6$ which arise from  spin-orbit mixing and rotation mixing, respectively,   with $^2\Sigma$ and $^2\Delta$ states.  
For MgF this mixing is substantially smaller (using parameters from Ref.~\cite{Doppelbauer2022}, the parity dependent $g$-factor for the $J^\prime=1/2$ state is $(g^\prime_l-g_r^{e^\prime})/3 \approx p/6B = 2 \times 10^{-4}$).  At this level of accuracy, the totality of all seven Zeeman interactions must be considered. The remaining $g$-factors have magnitudes $10^{-3}$ to $10^{-4}$.  
Some of these $g$-factors can be estimated from other spectroscopic parameters in the pure precession limit, but such estimates are suspect for MgF as  the pure precession hypothesis does not accurately predict the observed $\Lambda$-doubling of the MgF $\text{A}^2\Pi_{1/2}$ state \cite{Doppelbauer2022}. 
Without precision Zeeman spectroscopy, we cannot at present time definitively say much about the MgF $\text{A}^2\Pi_{1/2}$ $g$-factor beyond $\vert g \vert \lesssim 10^{-3}$. 
The sign of the $g$-factor is currently not known but will be determined experimentally by the laser polarizations which successfully trap molecules.
For our simulations, we use $g = 0.001$ as a representative value.

To address each of the ground state hyperfine levels, we simulate each laser beam as having three or four frequency components, denoted as (1)--(4) in Fig.~\ref{fig:laser_cool_trans}.
Frequency components (1)--(3) are all red-detuned by an equal amount $\Delta$ from their respective $F=1\rightarrow F'$, $F=0\rightarrow F'$ and $F=2 \rightarrow F'$ transition.
Frequency component (4) is blue detuned by $2 \Gamma$ from  the upper $F=1\rightarrow F'$ transition.
The blue-detuned frequency component provides additional spatial confinement at the cost of less damping for faster moving molecules~\cite{TarbuttII2015}.
Each of the six MOT beams has the same frequency components.

Experimentally, the four frequencies will be generated by acousto-optic modulators and subsequently recombined with polarizing and non-polarizing beamsplitters.
As such, it is technically easiest to have two of the four beams have the same polarization.
The chosen polarizations for the beams along $\pm z$ are shown in Fig.~\ref{fig:laser_cool_trans}.
We use Ref.~\cite{TarbuttII2015} as a guide for choosing the optimal polarization configuration.

We model the equations of motion and the population in each level using a rate equation model~\cite{Tarbutt2015} implemented in {\tt pylcp}~\cite{Eckel2022, NISTDisclaimer}, a python package capable of simulating laser cooling with complicated geometries and level structures.
Rate equations are used to compute the population of all 16 Zeeman sublevels, indexed by $i$, of either the ground ${\rm X}^2 \Sigma_+ (N=1)$ or excited ${\rm A}^2 \Pi_{1/2} (J^\prime=1/2)$ manifolds, $N^{\rm X,A}_i$, in the presence of lasers indexed by $l$, through
\begin{eqnarray}
     \dot{N}^{\rm X}_i & = & \sum_{j,l} R_{ij,ln} (N^{\rm A}_j-N^{\rm X}_i) + \sum_j \Gamma_{ji} r_{ji} N^A_j\\
    \dot{N}^{\rm A}_i & = & \sum_{j,l} R_{ji,ln} (N^{\rm X}_j-N^{\rm A}_i) - \Gamma N^A_i
\end{eqnarray}
where $\Gamma$ is the total decay rate of the ${\rm A}^2 \Pi_{1/2} (J^\prime=1/2)$ and $\Gamma_{ji}$ is the decay rate from state $j$ to state $i$.
Here, $R_{ij,ln}$ is the optical pumping rate of frequency component $n$ of laser $l$ defined by
\begin{equation}
    \label{eq:pumping_rate}
    R_{ij,lm} = \frac{\Omega_{ij,lm}^2/\Gamma}{1+4\{\omega_{lm}(t) -[\omega_j(\mathbf{r})-\omega_i(\mathbf{r})] - \textbf{k}_l \cdot \textbf{v} \}^2/\Gamma^2}\,,
\end{equation}
where $\omega_{ln}(t)$ is the time-dependent frequency of component $n$ of laser $l$, $\hbar \omega_j(\mathbf{r})$ is the position-dependent, Zeeman-shifted energy of state $j$ in the $A$ manifold, $\hbar \omega_i(\mathbf{r})$ is the energy of state $i$ in the $X$ manifold, $\mathbf{k}_{l}$ is the wavevector of laser $l$, $\mathbf{v}$ is the velocity of the molecule,
\begin{equation}
	\label{eq:rabi_rate_rate_eq}
	\Omega_{ij,ln} = \frac{\Gamma}{2} (\mathbf{d}_{ij}\cdot \boldsymbol{\epsilon}'_l) \sqrt{2s_{ln}(\mathbf{r},t)}
\end{equation}
is the Rabi rate, $\mathbf{d}_{ij}$ is the transition dipole moment between states $i$ and $j$, $\boldsymbol{\epsilon}'_l$ is the polarization of laser $l$, $s_{ln}(\mathbf{r},t)=I_{ln}(\mathbf{r},t)/I_{\rm sat}$ is the saturation parameter of frequency component $m$ of laser $l$ at position $\mathbf{r}$ and time $t$, and $I_{ln}$ is the intensity of frequency component $n$ of laser $l$.
The average force on the molecule is given by
\begin{equation}
    \label{eq:force}
    \mathbf{f} = \sum_{l}\frac{\hbar \mathbf{k_l}}{2} \sum_{i,j} R_{ij,l} (N^A_j-N^X_i)\,.
\end{equation}
The equilibrium force is determined by setting $\dot{N}_i^{X,A}=0$, solving for the populations, and inserting the result into Eq.~\ref{eq:force}.
Because this rate equation approximates optical coherences as having constant values, various sub-Doppler heating and cooling effects will be missing from the simulation.
For the loading process at large $\mathbf{r}$, the Zeeman shift is sufficient to force the optical coherences to oscillate rapidly and the rate equation approximation will be valid.
For simulations near the center of the MOT, the rate equation will most likely underestimate the temperature and size of the molecular cloud because it neglects sub-Doppler heating.

A few notational comments are in order.
Because the six MOT beams have identical intensities and frequency components, we drop the superfluous $l$ index unless necessary.
We specify $\omega_{m}$ in Eq.~\ref{eq:pumping_rate} in terms of the detuning $\Delta_{m}=\omega_{m}(t) -[\omega_{F}-\omega_{F'}] $ relative to the zero-field energies $\hbar\omega_F$ and $\hbar\omega_{F'}$ that the frequency component $m$ is intended to drive.
Here, $\hbar\omega_{F'}$ is always the unperturbed energy of the $\ket{{\rm A}^2\Pi_{1/2},J^\prime=1/2}$ state.
For frequency components $m=1$ and $m=2$, $\hbar\omega_F$ is the energy of the lower $F=1$ and $F=0$, respectively.
For frequency components $m=3$ and $m=4$, $\hbar\omega_F$ is the mean energy of the upper $F=1$ and $F=2$ states.
Finally, we denote the saturation parameters of the four frequency components as a vector $\mathbf{s} = (s_1,s_2,s_3,s_4)$.
For Gaussian beams, $\mathbf{s}$ denotes the maximum saturation parameters at $\mathbf{r}=0$.

For the results of Sec.~\ref{sec:static_MOT}, we first determine the equilibrium force as a function of $v$ and $x$ and then evolve the motion of the molecule using that force.
We have verified that, to much better than the expected accuracy of the simulations, our approach agrees with the result if the time evolution of both the motion and the internal state populations are calculated using the full rate equation model. 
This simplification greatly reduces the computational complexity, reducing the number of differential equations from 18 (16 internal states, 1 velocity, and 1 position) to 2 (1 velocity and 1 position), albeit through a complicated force versus position and velocity profile.

For the results of Sec.~\ref{sec:chirped_MOT}, we calculate the equilibrium force not just as a function of $v$ and $x$, but also as a function of the common detuning $\Delta$.
We evolve the population of internal states and motion of the molecule through this three-dimensional force profile, given a function $\Delta(t)$.

For Sec.~\ref{sec:stability}, we include spontaneous emission effects by including random momentum kicks with a probability that is proportional to the excited state populations.
For more details, see Ref.~\cite{Eckel2022}.
Effects of momentum diffusion due to stimulated emission are neglected.

\section{Capture into a static MOT}
\label{sec:static_MOT}

Consider the properties and capture process of a static MOT, which has constant $\Delta(t)=\Delta_0$.
While this process has been discussed in the literature before \cite{Haubrich1993,Eckel2022}, it is nonetheless illustrative and will help motivate our choices for MgF.
For this discussion, let us first consider the capture process for a $^{87}$Rb type-I MOT with infinite plane wave beams arranged in the geometry described above \footnote{Here, we simulate the $F=2\rightarrow F'=3$ transition and neglect the off resonant $F=2\rightarrow F'=2$ scattering that requires the $F=1\rightarrow F'=2$ repump.}.
We use ``natural'' units of the MOT, where velocities are measured in terms of $\Gamma/k$ and positions are measured in terms $\hbar \Gamma/\mu_B B'$; that is, velocity and position are measured by the number of natural linewidths which equal the  Doppler and Zeeman shifts, respectively.
For $^{87}$Rb with a $B'=2$~mT/cm $B$-field gradient, $\hbar \Gamma/\mu_B B'\approx 2$~mm and $\Gamma/k\approx 46$~m/s.

The calculated force profile driven by a single frequency component labeled $m=0$ with $s_{0}=I_{0}/I_{\rm sat} = 2.5$ and $\Delta_0=-\Gamma$ is shown in Fig~\ref{fig:explainer_fig}(a).
The force profile consists of three ``bands'' of both positive and negative force, which correspond to when one or more polarization components of the lasers are Doppler and/or Zeeman shifted into resonance.
According to Eq.~\ref{eq:pumping_rate}, and using $|\mathbf{k}\cdot \mathbf{v}| = \sqrt{2} k v$ for our geometry, these resonances occur when 
\[
    \frac{\Delta_0}{\Gamma} \pm \frac{k v}{\sqrt{2}} + \epsilon_i\frac{\mu_B B'}{2\hbar \Gamma} x = 0\,,
\]
where $\epsilon_i = -1, 0, 1$ for the $\sigma^-$, $\pi$ and $\sigma^+$ components of the light projected onto the $x$ axis.
Here, we have inserted the approximate differential Zeeman shift for an alkali of $\omega_i-\omega_j = \mu_B B' x/2\hbar$.
The $+$($-$) sign occurs when the beams are mostly counter-propagating (co-propagating) to the incoming atoms.
The dominant $\sigma^+$ from the predominantly counter-propagating beams and dominant $\sigma^-$ from the predominantly co-propagating beams form the $-1/\sqrt{2}$ slope positive and negative forces, respectively.
Likewise, the weak $\sigma^-$ component from the predominantly counter-propagating beams and the weak $\sigma^+$ from the predominantly co-propagating beams form the $+1/\sqrt{2}$ slope positive and negative forces, respectively.
Finally, the $\pi$ components from the beams form the zero slope force curves.

The calculated trajectories through the force profile, shown Fig~\ref{fig:explainer_fig}(a), reveal the capture process in a $^{87}$Rb MOT.
Atoms enter the MOT with $x<0$ and $v>0$.
For $0<v\lesssim 3 \Gamma/k$, atoms are slowed and stopped by the $\pi$ component.
While these slowest atoms do not reach the origin after the 20~ms integration time, a small off-resonant spatial force from the predominantly counter-propagating beams will eventually push these atoms to $x=0$.
Faster atoms with $3 \Gamma/k\lesssim v \lesssim 8 \Gamma/k$ initially experience a boost from the predominantly co-propagating beams, but then fall onto a nearly common trajectory of being slowed and trapped by the Zeeman- and Doppler-shifted predominantly counter-propagating beams.
These trajectories terminate at $v=0$ and $x=0$, indicating successful capture.
For $v \gtrsim 8 \Gamma/k$, the boost from the predominantly co-propagating beams is too large to be overcome by the counter-propagating beams, and the atoms evade capture.

\begin{figure*}
    \centering
    \includegraphics{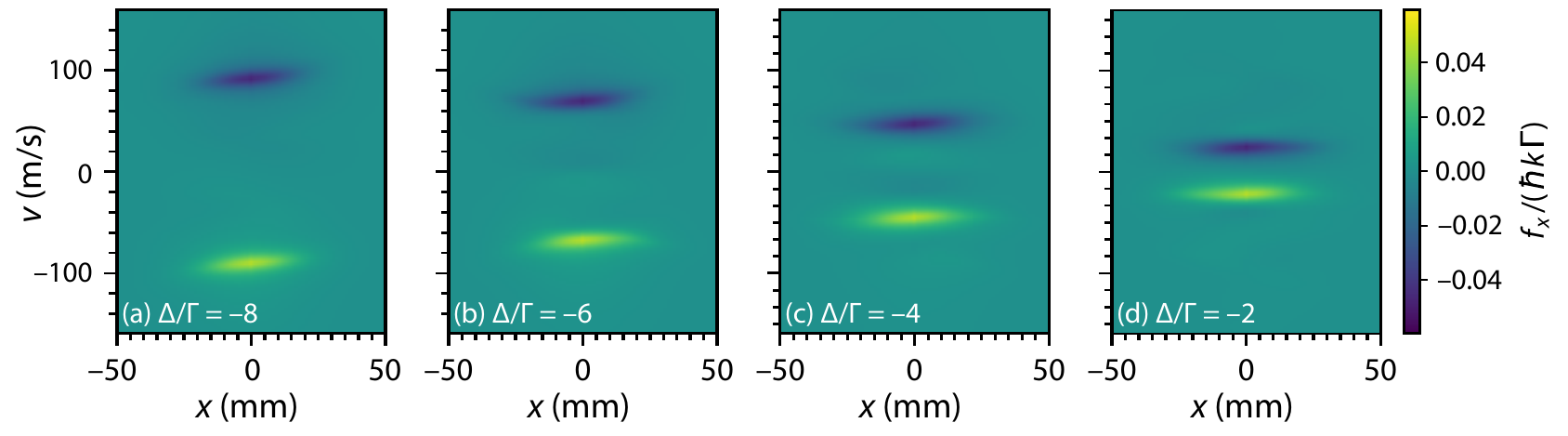}
    \caption{Force profiles of a MgF MOT using three frequency components, elliptical lasers beams with $w_{xy}=17.5$~mm and $w_z=10$~mm, and using a total laser power of 1~W. Equilibrium force profiles are shown at select detunings to illustrate the time-varying force profile in a chirped MOT.  The detunings from panels (a) to (d)  are $\Delta=-8\Gamma$, $\Delta=-6\Gamma$, $\Delta=-4\Gamma$, and $\Delta=-2\Gamma$, respectively.  As the detuning becomes less negative,  the force maximizes at progressively slower velocities while the magnitude of the force remains roughly constant.}
    %\caption{The solutions of the rate equations encode both the damping and spatial forces into color plots: the dark band representing a negative force and the yellow band representing a positive force with each able to address velocities with the opposite sign. Trajectories of MgF particles evolved for 20ms with different initial velocities overlaid onto their corresponding force profiles}
    \label{fig:force_profiles_with_detuning}
\end{figure*}

We now contrast the capture process of a Rb atom to that of MgF.
Let us first consider a static MOT with infinite plane wave beams containing frequency components(1)--(3) from Fig.~\ref{fig:laser_cool_trans}: the transitions $F=1\rightarrow F'$, $F=0\rightarrow F'$ and $F=1,2\rightarrow F'$ are all addressed by a frequency component with $\Delta = -\Gamma$, labeled by (1)-(3) in Fig.~\ref{fig:laser_cool_trans}.
Here, the saturation parameters of the frequency components are chosen to be $\mathbf{s} = (1.45, 1.45, 2.89, 0)$.
This $\mathbf{s}$ correspond to having relative saturation parameters $\tilde{\mathbf{s}} = \mathbf{s}/(\sum_n  s_n) = (0.25, 0.25, 0.50, 0)$ in a ``prototypical'' 1~W Gaussian beam with waists $w_{xy}$ and $w_z$ equal to 17.5 and 10~mm respectively.
For MgF, the natural length and velocity scales for the MOT are $\hbar \Gamma/\mu_B B'= 7.48(8)$~mm with $B'=2$~mT/cm and $\Gamma/k= 7.53(8)$~m/s.

Compared to $^{87}$Rb, there are three significant differences.
First, the maximum force is much lower in the MgF MOT because the type-II level structure requires constant repumping of states which are not coupled to laser beams which provide a restoring force.
With $n_g=12$ ground states and $n_e=4$ excited states, our anticipated maximum scattering rate is no greater than $R_{\rm max} = \Gamma/4$.
Indeed, the maximum scattering rate (not shown) in Fig.~\ref{fig:explainer_fig}(b) and (c) is approximately $R_{\rm max}/2$ at $v=\pm\sqrt{2}(\Gamma/ k)$ and $x=0$, due to having $s_n\approx 1$ for all transitions.
Assuming that all the scattering is due to counter-propagating beams, one might expect that the maximum force to be $R_{\rm max} \hbar k/\sqrt{2}$, where the factor of $\sqrt{2}$ comes from the projection of the counter-propagating MOT beams onto the axis.
We instead observe that the force is reduced to approximately $R_{\rm max} \hbar k/2\approx 0.05\times \hbar k \Gamma$, because roughly 30~\% of the photon scatters are from the $\pm \hat{z}$ beams and 10~\% are from the co-propagating beams at $v=\sqrt{2}(\Gamma/ k)$ and $x=0$.

Second, because of the small excited state $g$-factor and the presence of dark states on the type-II transitions, the MgF MOT has no appreciable force outside of $|x|>5(\hbar \Gamma/\mu_B B')$ and virtually no slope to the force.
This greatly reduces the capture velocity from $v \lesssim  10(\Gamma/k)$ for Rb to $v_c \approx  4(\Gamma/k)$ for MgF.

Third, MgF molecules with initial velocities $v\lesssim 4(\Gamma/k)$ failed to arrive at $x=0$ within the maximum integration time of 20~ms.
This is due to the reduced trapping force in a type-II MOT.
This reduction in spatial trapping force is further compounded by the fact that the unresolved $F=1$ and $F=2$ states are driven by the same laser, which has the correct polarization to trap $F=2$ but, necessarily, the incorrect polarization to trap $F=1$ \cite{Tarbutt2015}.

In Fig.~\ref{fig:explainer_fig}(c), we attempt to increase the spatial confinement by adding frequency component (4) shown in Fig.~\ref{fig:laser_cool_trans}, which is blue-detuned from the unresolved $F=1,2\rightarrow F'$ transition~\cite{Tarbutt2015,Jarvis2018,Burau2023}.
The saturation parameters are chosen to be $\mathbf{s} = (1.45, 1.45, 2.17, 0.72)$, corresponding to $\tilde{\mathbf{s}} = (0.25,0.25,0.375,0.125)$ for our prototypical beam parameters.
While frequency component (4) again has the correct polarization to trap $F=2$ but the incorrect polarization to trap $F=1$.
Nonetheless, with the additional trapping force, molecules entering the MOT with $v\lesssim 4(\Gamma/k)$ reach the origin within 20~ms.
The presence of this component adds a slight acceleration at large, negative $x$ that causes $v = 4.2(\Gamma/k)$ to just barely be trapped.

We now consider the effect of overall detuning on the force profiles for MgF given more experimentally realistic elliptical Gaussian beam profiles, as described in Sec.~\ref{sec:model}.
Fig.~\ref{fig:force_profiles_with_detuning} show the force profiles, without trajectories, for four detunings $\Delta_n/\Gamma$ for the three-frequency-component configuration with $\mathbf{s} = (1.45, 1.45, 2.89, 0)$.

The force profiles reveal well-separated positive and negative force regions, with extrema at $x=0$ and $v=\pm\sqrt{2} (|\Delta|/k)$.
Each region resembles a ``boat''--a two-dimensional Gaussian with a rough $1/e^2$ half-width of $\Gamma/k\approx 7.5$~m/s in $v$ and $\sqrt{2}w_{xy}\approx 25$~mm--floating in sea of zero force.
The shape in the $x$ direction is a convolution of the Gaussian beam profile and the shape seen in Fig.~\ref{fig:explainer_fig}(b) caused by Zeeman dark states.
Thus, increasing the beam size beyond $\sqrt{2} w_{xy} \gtrsim 4 (\mu_B B'/\hbar \Gamma)$ or, equivalently, $w_{xy} \gtrsim 21~$~mm will generally not result in a larger spatial extent of the force.

Trajectories through the force profiles Fig.~\ref{fig:force_profiles_with_detuning} (not shown) are generally not trapped.
Consider Fig.~3(a).
A molecule entering from the left with velocity $v<50$~m/s ($6.7\times\Gamma/k$) or $v>120$~m/s ($6.7\times\Gamma/k$) will not be affected by the isolated negative force centered at $v=\sqrt{2} \times 8 \Gamma/k$ and will fly straight through the MOT.
Likewise, molecules with $50<v<120$~m/s will be slowed but will not be trapped in the MOT.
Thus, we see that static-detuning force profiles lack a smoothly-connected decelerating force from high velocity to zero velocity, an  essential feature of an alkali MOT.

\section{Capture in a frequency-chirped MOT}
\label{sec:chirped_MOT}

To engineer a smoothly-connected force from large $v$ to small $v$, we ramp $\Delta$ from large to small negative values over a duration $\tau$.
Note that the maximum force and therefore maximum deceleration is roughly constant with $\Delta$ (see Fig.~\ref{fig:force_profiles_with_detuning}).
Under constant deceleration, the velocity decreases linearly with time, which requires a linear ramp of $\Delta$ to maintain Doppler-shifted resonance, i.e.,
\begin{equation}
    \Delta_m(t) = \left\{
    \begin{array}{lr} 
    \Delta_{\rm I} + \frac{\Delta_{\rm F}-\Delta_{\rm I}}{\tau} t & 0< t <\tau \\
    \Delta_{\rm F} & t>\tau
    \end{array}
    \right.\,,
    \label{eq:freq_chirp}
\end{equation}
for $m=1,2,3$.
Choosing the parameters $\Delta_{\rm I}$, $\Delta_{\rm F}$ and $\tau$ are of utmost importance.

To make an initial estimate of $\Delta_{\rm I}$, $\Delta_{\rm F}$ and $\tau$, let us consider a simple model where a constant force $f$ is applied over a distance $d$.
The maximum velocity that can be stopped across that distance is $v_{\rm c}=\sqrt{2 f d/m}$, which will occur in a time $\tau =v_{\rm c}/(f/m)$ where $m$ is the mass of the molecule.
Using roughly $f \approx 0.03\times \hbar k \Gamma$ from Fig.~\ref{fig:force_profiles_with_detuning} and $d\approx 30$~mm, we find $v_{\rm c}\approx 80$~m/s and $\tau=0.8$~ms.

We simulate capture into such a frequency-chirped MOT.
Our chirped MOT begins in the three-frequency component configuration with a common detuning of $\Delta_{\rm I} = -8\Gamma$ and $\mathbf{s}=(1.45, 1.45, 2.89, 0)$, which could potentially address all velocity classes up $\sqrt{2}\times 8\Gamma/k \approx 84$~m/s.
At $t=\tau=1$~ms, the frequency chirp ends at $\Delta_{\rm F}=-\Gamma$.
We then instantaneously switch to the four-frequency component configuration with $\mathbf{s} = (1.45, 1.45, 2.17, 0.72)$ to enhance our spatial confinement, as observed in Sec.~\ref{sec:static_MOT}.
These two sets of saturation parameters correspond to the peak saturation parameters of a 1~W in a Gaussian beam with $w_{xy}$ and $w_z$ equal to 17.5 and 10~mm, respectively, and relative $\tilde{\mathbf{s}}=(0.25,0.25,0.5,0)$ for the three-frequency component case and $\tilde{\mathbf{s}}=(0.25,0.25,0.375,0.125)$ for the four-frequency component case.
These choices are the same as in Sec.~\ref{sec:static_MOT}.

\begin{figure}
    \centering
    \includegraphics{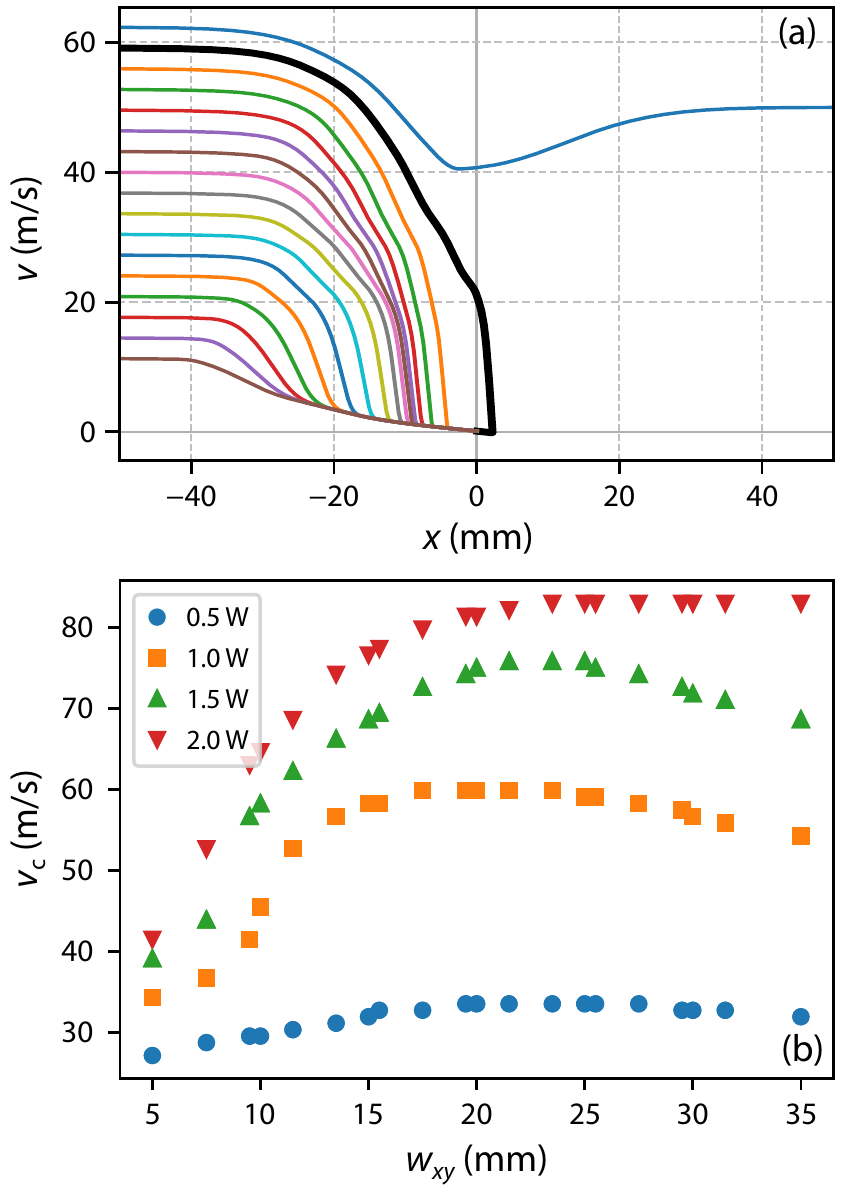}
    \caption{
    (a) Classical phase space trajectories of MgF molecules with various initial velocities in a frequency-chirped MOT with $1/e^2$ beam radii of $w_{xy}=17.5$~mm and $w_z=10$~mm and axial magnetic field gradient $B'=2$~mT/cm, using the frequency chirp of Eq.~(\ref{eq:freq_chirp}) with $\Delta_0=-8\Gamma$, $\Delta_1=-\Gamma$, $\tau=1$~ms, and total laser power of 1~W.
    The thick, black curve shows the trajectory with the largest initial velocity that is captured.
    (b) Capture velocity $v_{\rm c}$ for a frequency-chirped MOT vs. beam waist $w_{xy}$ for various total powers.
    Other parameters are the same as in (a).
    }
    \label{fig:capture_vel_chirp}
\end{figure}

Figure~\ref{fig:capture_vel_chirp}(a) shows the resulting trajectories of molecules through classical phase space.
The initial position of the molecules is $x_0 = -50$~mm, such that they start far from the position of maximum force (see Fig.~\ref{fig:force_profiles_with_detuning}).
In this configuration, the maximum velocity class captured is $v_{\rm c}=7.5(\Gamma/k)\approx 57$~m/s. This $v_{\rm c}$ is about double that of the $v_{\rm c}$ observed in the static MOT of Fig.~\ref{fig:explainer_fig}(c) and approaches the velocity observed in two-stage CBGB sources~\cite{Hutzler2012, Lu2011, Hemmerling2014,pattersonBrightGuidedMolecular2007}.

We also study the dependence of the capture velocity both on laser power and $w_{xy}$. Fig.~\ref{fig:capture_vel_chirp}(b) shows the results. The maximum capture velocity observed in our simulations, with 2~W of laser power and identical $\tilde{\mathbf{s}}$ to those above, is 80~m/s, or $10.5\times(\Gamma/k)$.
We observe two regimes: one with $v_c<45$~m/s and a second with $v_c>45$~m/s. Given that the static MOT with similar parameters in Fig.~\ref{fig:explainer_fig}(b) showed a capture velocity of $v_c\approx 4\Gamma/k\approx 32$~m/s, we conclude that these two regimes denote ineffective and effective chirped slowing.

To understand the ineffective chirped slowing regime, observe that for a given starting position, molecules require some initial evolution time to encounter the small-extent spatial force of the MOT.
For example, a molecule moving at 50~m/s is unperturbed for at least 0.8~ms before encountering a force from a $w_{xy}\leq 10$~mm MOT beam.
By this time, the velocity at which the slowing force is maximal is at $v=2.7\times(\Gamma/k)\approx 20$~m/s, well below the 50~m/s initial velocity.
These molecules simply missed the boat.
Likewise, while we calculate the capture velocity for molecules starting at the same position; in reality, there will be a distribution of starting positions, and some of the  molecules near the ends of that distribution may also miss the boat.
These complications highlight the well-known problem of optimizing frequency-chirp slowing for both starting position {\it and} velocity \cite{Truppe2017b}.

With a total beam power of 0.5~W and chirp duration $\tau = 1$~ms, chirped slowing is predicted to be ineffective for any $w_{xy}$.
The slowing force exerted by the chirped beams at 0.5~W is somewhat weaker at roughly $f \approx 0.015 \times \hbar k \Gamma$, consequently requiring a longer chirp of at least $\tau = 1.6$~ms to effectively decelerate the molecules than the $\tau = 1$~ms rate simulated in Fig.~\ref{fig:capture_vel_chirp}.
In keeping with our analogy, while these molecules may have caught the boat, the boat was moving too fast for the molecules to remain on.

In the second, effective chirped slowing regime, the MOT beams are both sufficiently large and powerful.
The capture velocity in this regime initially increases with increasing $w_{xy}$, reaches a maximum, and subsequently slowly decreases.
To understand this shape, let us approximate $v_{\rm c}\approx \sqrt{2 f d/m}$, where $f$ is a constant force applied over an effective distance $d$.
If the transitions were unsaturated and the force at large distances not attenuated by Zeeman substates being tuned out of resonance, $v_{\rm c}$ would be independent of $w_{xy}$, because $f \propto I \propto 1/w_{xy}$ and $d\propto w_{xy}$.
At small $w_{xy}$, the transitions are somewhat saturated, and $f$ decreases more slowly than $1/w_{xy}$ with increasing $w_{xy}$.
Coupled with the $d\propto w_{xy}$, this modified dependence of $f$ implies increasing $v_{\rm c}$ with increasing $w_{xy}$.
At large $w_{xy}$, $d$ no longer scales directly with $w_{xy}$, but instead is set by a convolution of the Gaussian beam shape and the attenuation of the force at $x\gtrsim 4(\hbar \Gamma/\mu_B B')$ due to Zeeman sublevels being shifted out of resonance, as seen in  Fig.~\ref{fig:explainer_fig}(b)-(c)].
Note that, for $B'=2$~mT/cm, this convolution means that $d$ no longer grows linearly with $w_{xy}$ for $w_{xy}\gtrsim 20$~mm.
At these large $w_{xy}$, $d$ increases slower than linearly with $w_{xy}$ while $f\propto 1/w_{xy}$, and thus $v_{\rm c}$ decreases with increasing $w_{xy}$.
This could potentially be improved by reducing $B'$.

\section{Stability of MOT during chirp}
\label{sec:stability}

\begin{figure}
    \centering
    \includegraphics{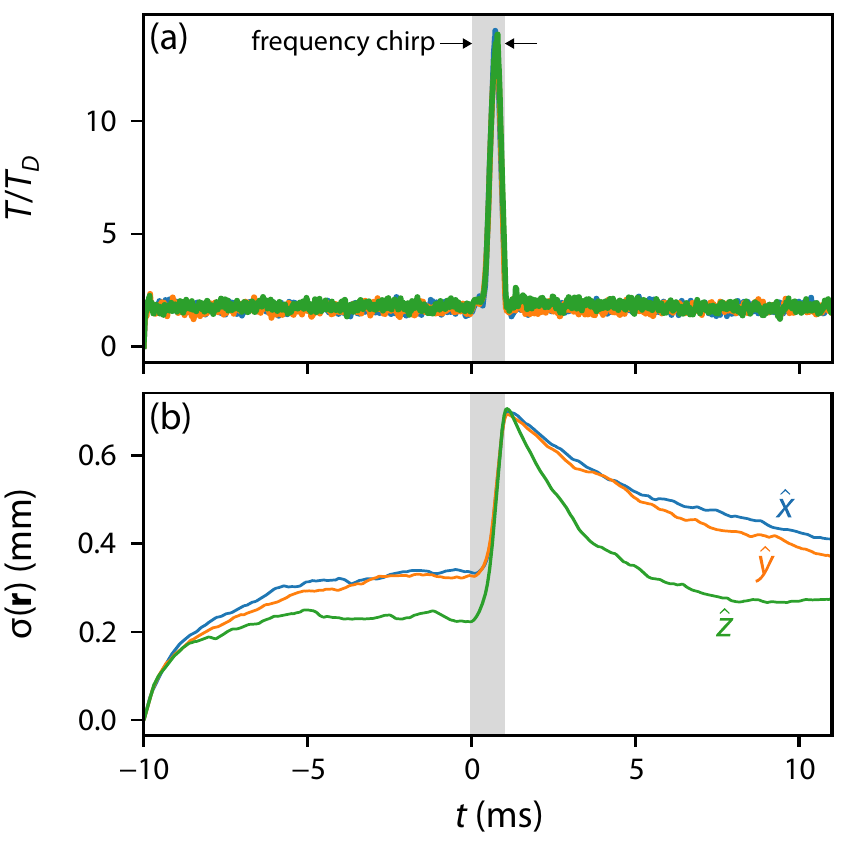}
    \caption{
    (a) Simulated temperature $T$, independently calculated across all three spatial dimensions $x$ (blue), $y$ (orange) and $z$ (green) and normalized to the Doppler temperature $T_D$, of a MOT before, during and after the frequency chirp (gray band) of 320 simulated molecule trajectories.
    (b) Size $\sigma(\mathbf{r})$ of the MOT.}
    \label{fig:std_time}
\end{figure}

We now turn to the stability of the MOT during the frequency chirp.
In order to load multiple molecular pulses from a CBGB source into a chirped MOT, we need to verify the stability of any previously loaded molecules in the MOT when a subsequent frequency chirp occurs.
Of greatest concern during the frequency chirp is the heating that may occur.
In particular, as illustrated in Fig.~\ref{fig:laser_cool_trans}, as the detuning is changed according to Eq.~(\ref{eq:freq_chirp}), the frequency components (1) and (2), which are intended to address the lower $F=1\rightarrow F'$ and $F=0\rightarrow F'$ transitions of the moving molecules, respectively, will incidentally sweep through resonance with the $F=0\rightarrow F'$ and $F=1,2\rightarrow F'$ transitions of molecules already in the MOT, respectively.
The first of these resonances occurs at $\Delta_m\approx -6\Gamma$; the second occurs at $\Delta_m\approx -5.5\Gamma$.
Thus, we anticipate that the MOT will heat during the frequency chirp.

We simulate the heating by solving for the motion of 320 molecules in a chirped MOT using Eq.~(\ref{eq:freq_chirp}) with the same parameters as in Sec.~\ref{sec:chirped_MOT}.
In these simulations, we use plane waves for computational simplicity.
The plane waves have three frequency components with $\mathbf{s}=(1.45, 1.45, 2.89, 0)$ during the chirp and four frequency components with $\mathbf{s} = (1.45, 1.45, 0.72, 2.17)$ before and after the chirp.
As in Secs.~\ref{sec:static_MOT} and \ref{sec:chirped_MOT}, these saturation parameters correspond to the peak saturation parameters of our prototypical 1~W in a Gaussian beam with $w_{xy}$ and $w_z$ equal to 17.5 and 10~mm, respectively, and relative $\tilde{\mathbf{s}}=(0.25,0.25,0.5,0)$ for the three-frequency component case and $\tilde{\mathbf{s}}=(0.25,0.25,0.375,0.125)$ for the four-frequency component case.
The simulated MOT is spatially compact with typical size $\sigma_r<1$~mm, thus using Gaussian beams with a $1/e^2$ radius $>10$~mm induces at most a small error. Unlike simulations of the capture process, we compute motion along all three spatial dimensions and include momentum diffusion due to spontaneous emission.

We initialize the particles at $t=-10$~ms with $\mathbf{v}=0$ and $\mathbf{r}=0$.
This initial condition is chosen for two reasons:
(1) we do not know {\it a priori} the size and temperature of the simulated MOT, and
(2) by observing evolution of the MOT toward equilibrium, we can extract relaxation times independent of the frequency chirp.
The frequency chirp begins at $t=0$~ms and lasts until $t=1$~ms.
The simulation continues with four fixed frequencies until $t=11$~ms to understand the trends back toward equilibrium.
%Approximately 11~h is required to compute a single molecule trajectory on a standard desktop computer.\DSB{This can go. Something to tell a reviewer if they ask about the number of trajectories.}
The state population, position, and velocity of each molecule are recorded at 2.1~$\mu$s intervals.
%We then compute the mean and standard deviation of each component of the velocity and position across the distribution of 273 molecules.
Temperatures at each time are assigned using the relation $\sigma_{v_i}^2 = k_B T/m$, where $k_B$ is the Boltzmann constant and $\sigma_{v_i}$ is the standard deviation of the velocity $v_i$ along $i=x,y,z$.
%from the
%The standard deviation of the velocity $\sigma_{v_i}$, where $i=x,y,z$, is related to the temperature through $\sigma_{v_i}^2 = k_B T/m$, where $k_B$ is the Boltzmann constant.

The size and temperature of the simulated MOT is shown in Fig.~\ref{fig:std_time}.
Before the chirp, the MOT temperature settles to about $1.7~ T_D$, where $T_D=\hbar \Gamma/2 k_B$ is the Doppler temperature.
This temperature is lower than those typically observed in molecular MOTs \cite{Norrgard2016,andereggRadioFrequencyMagnetoOptical2017,williamsCharacteristicsMagnetoopticalTrap2017} because our rate equation model lacks both momentum diffusion caused by stimulated emission and sub-Doppler heating.
As anticipated, we see a rapid increase in the temperature of the MOT during the frequency chirp, rising from $1.7~T_D$ to about $14~ T_D$.
After the the chirp, however, the MOT returns to its equilibrium temperature within 100~$\mu$s.
The mean velocity (not shown) remains zero for all $t$. 

The measured size of the MOT is more complicated.
Before the chirp, the MOT trends slowly towards its equilibrium size of roughly 0.4~mm $e^{-1/2}$ radius.
After the chirp, the MOT has expanded to roughly $0.7$~mm in size due to heating, but slowly contracts back toward equilibrium, faster along $z$ with the stronger magnetic field gradient, and slower along $x$ and $y$. 
The mean position (not shown) remains zero for all $t$.

Critically, no molecule in our simulation appears to be lost, that is, gaining a velocity that could not be subsequently damped.
Experimentally, the MOT will most likely have an initial $T/T_D \approx 4$ and $\sigma \approx 1$~mm.
Yet, assuming proportional heating to $T/T_D \approx 28$, the MOT would only expand to $\sigma_z \approx 3$~mm.
While approaching the $w_z=10$~mm, the distance at which the spatial component of the force is seriously diminished, it is still comfortably below that limit.
Likewise, based on Figs.~\ref{fig:explainer_fig}(c), damping forces exist for $|v| \approx (\Gamma/k)$, which should effectively cool a molecular cloud with a temperature as high as  $T/T_D \approx 600$.
Thus we anticipate that most molecules remain trapped in the MOT even under this pessimistic scenario.

\section{Conclusion}
\label{sec:conclusion}
% , a consequence of the physical properties of the molecule

We have proposed and theoretically investigated a frequency-chirped MOT for laser-coolable lightweight molecules like MgF.
The frequency-chirped MOT has a maximum capture velocity for MgF of about 80~m/s, which is commensurate with typical molecular beam velocities observed using a two-stage cryogenic buffer gas beam source~\cite{Hutzler2012, Lu2011, Hemmerling2014}.
Compared to standard frequency-chirped slowing, our frequency-chirped MOT has advantages and disadvantages.

The biggest disadvantage is that the force is reduced by a factor of $\cos{\theta}$, where $\theta$ is the projection of the laser beam's $k$-vector on the molecular beam axis.
Thus, each photon scattered is less effective in slowing than in standard frequency-chirped slowing.

The biggest advantage is the potential for loading of multiple molecular pulses from a CBGB source, which could greatly increase the number of captured molecules.
We have shown that the molecules in the MOT should not be lost during the frequency chirp.
This contrasts to traditional chirped slowing, where a single slowing beam intersects the MOT causing a resonant, directed force during the frequency chirp that ejects molecules from the MOT.
In this limit, the equilibrium population in the MOT will be determined by the number of molecules captured per CBGB pulse, the frequency of pulses, and the %vacuum-limited 
lifetime of the MOT.
We note that the lifetime of the MOT must be comparable to or longer than the duration between CBGB pulses in order to realize this gain, and likely requires tuning of the MOT beam parameters during the time between capture of one pulse and the start of the next \cite{andereggRadioFrequencyMagnetoOptical2017,williamsCharacteristicsMagnetoopticalTrap2017} beyond the simple parameters simulated here.
Further optimization of such parameters will be the subject of future theoretical and experimental work.

The proposed technique with likely work for light molecules such as MgF, BeF, BeH, BH, and AlF.
For heavier molecules like CaF, SrF, YbO, and YbF, the stopping distances are much larger than typical MOT beam sizes.
One intriguing possibility for such heavy molecules is to combine the chirped-MOT  with the chirped \cite{Truppe2017b} or white-light \cite{Barry2012} slowing typically used to load fixed-frequency MOTs.  In such a configuration, laser slowing could enable loading of a chirped-MOT while being sufficiently far from resonance to not perturb trapped molecules.  Such a hybrid technique may then enable loading multiple pulses of heavier molecules into a MOT but requires further investigation. 

\section*{Acknowledgments}
K. Rodriguez thanks Benjamin Goldweber and Jeff Miller for programming assistance.
The authors thank David La Mantia for useful discussions and Eric Shirley and Jabez McClelland for a thorough reading of the manuscript.
This work was supported by NIST.

\bibliography{main}

\begin{thebibliography}{43}
\bibitem [{NIS()}]{NISTDisclaimer}%
  \BibitemOpen
  \href@noop {} {}\bibinfo {note} {Certain equipment, instruments, software, or
  materials are identified in this paper in order to specify the experimental
  procedure adequately. Such identification is not intended to imply
  recommendation or endorsement of any product or service by NIST, nor is it
  intended to imply that the materials or equipment identified are necessarily
  the best available for the purpose.}\BibitemShut {Stop}%
\end{thebibliography}

\end{document}